\documentclass[sigconf]{acmart}

\copyrightyear{2018}

\usepackage{booktabs} 

\title{Siamese Cookie Embedding Networks \\ for Cross-Device User Matching}

\usepackage{amsmath}
\usepackage{amssymb}
\usepackage{tikz}
\usetikzlibrary{fpu}
\usepackage{xspace}
\usepackage{subfig}

\newcommand{\simcnn}{\emph{SCEmNet}\xspace}
\newcommand{\jsimcnn}{\emph{JSCEmNet}\xspace}

\tikzset{
  font={\fontsize{18pt}{18}\selectfont}}

\author{Ugo Tanielian}
\email{u.tanielian@criteo.com}
\affiliation{Criteo Research}
\author{Anne-Marie Tousch}
\email{am.tousch@criteo.com}
\affiliation{Criteo Research}
\author{Flavian Vasile}
\email{f.vasile@criteo.com}
\affiliation{Criteo Research}

\begin{document}

\maketitle

\section*{Abstract}

Over the last decade, the number of devices per person has increased substantially. This poses a challenge for cookie-based personalization applications, such as online search and advertising, as it narrows the personalization signal to a single device environment. A key task is to find which cookies belong to the same person to recover a complete cross-device user journey. Recent work on the topic has shown the benefits of using unsupervised embeddings learned on user event sequences. In this paper, we extend this approach to a supervised setting and  introduce the \emph{Siamese Cookie Embedding Network} (\simcnn), a siamese convolutional architecture that leverages the multi-modal aspect of sequences, and show significant improvement over the state-of-the-art.

\section{Introduction}
The cross-device matching task has been addressed recently in the literature, most notably within the scope of the ICDM 2015\footnote{ICDM 2015:~\url{https://www.kaggle.com/c/icdm-2015-drawbridge-cross-device-connections}} challenge and the CIKM Cup 2016\footnote{CIKM 2016:~\url{http://cikm2016.cs.iupui.edu/cikm-cup/}}. 

The winners of the CIKM Cup \citep{Phan2016, Phan2017} proposed using neural features to describe sequences of URL tokens. Their approach is based on a hierarchical \emph{doc2vec} model~\citep{DaiOL15}. However, in real world settings, \emph{doc2vec} suffers from several drawbacks: first, \emph{doc2vec} generates unsupervised sequence embeddings that are not specialized for the final task; secondly, \emph{doc2vec} generates embeddings only for sequences available at training time.

In our approach, we propose to use the \textit{TextCNN} structure as in \citet{Kim2014} in a siamese network to learn cookie similarities with different types of sequences. Using a supervised approach allows us to reach better performance than the state-of-the-art. Furthermore, our approach is able to match cookies unseen at training time. 

Our main contribution is a new architecture for the problem of cookie matching, the \simcnn structure, and its wide\&deep counterpart (\jsimcnn) that address the shortcomings of previous embedded-based works.

\section{Our approach}\label{sec:approach}

\subsection{Context}
We evaluate our approach on a cross-device matching data set that was released for the CIKM Cup in 2016. The description of the data set can be found in \citet{Phan2017}. The data set is composed of about 339k cookies and more than 506k ground truth pairs in training. The task is to find the 215k true pairs linking the users in test. For each cookie, the data set contains a sequence of events, where an event is a pair (URL, timestamp). The evaluation metrics of the challenge are precision, recall, and F1-score.

For a given set of cookies, the aim is then to compute a list of pairs of matched users. Since evaluating all possible pairs between $n$ cookies results in $O(n^2)$ operations, a classical 2-step approach is to have a candidate pair generation followed by pairwise ranking trained on ground truth data. 

In the first step, we use a cookie representation similar to~\citet{Phan2017} with k-nearest-neighbors to generate the candidate pairs. For the ranking step, we first implemented a baseline ranking algorithm. For each of the candidate pairs, we generate a set of \textit{Baseline features}: some features are based on URL sequences such as TF, TF-IDF similarities, term-matching features and \emph{doc2vec} embedding similarities, while other features are time-related features. For better results, we introduce our supervised \simcnn architecture.

\subsection{\simcnn Architecture}
\begin{figure}
\subfloat[\emph{SeqCNN} module]{\label{fig:cnnmodule}
\resizebox{\columnwidth}{!}{
\input{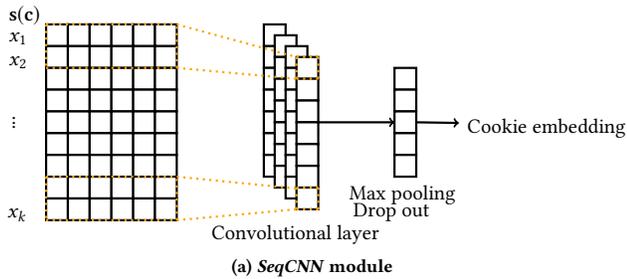}
}}
\hspace{5pt}
\subfloat[\simcnn module]{\label{fig:multicnn}
\resizebox{\columnwidth}{!}{
\input{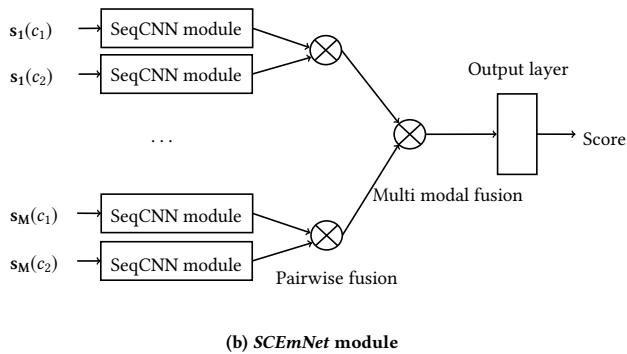}
}}
\caption{\simcnn architecture: fusionning several pair embeddings to get a richer one.}
\label{multSimcnn}
\end{figure}

Our similarity learning architecture is based on convolutional networks and is illustrated in Figure~\ref{multSimcnn}. We rely on the assumption that each cookie $\mathbf{c}$ is associated with $M$ sequences ($\mathbf{s_1(c),...,s_M(c)}$) of fixed size. For a given token modality, the sequence of tokens is first processed through a \emph{SeqCNN} module, similar to the \textit{TextCNN} architecture introduced in~\citet{Kim2014}: given an input sequence of embedded tokens, a convolutional network with one convolution, max-pooling and dropout layer outputs a \emph{cookie embedding}. We use this module in a siamese network with a pairwise fusion layer to transform the two cookie embeddings into a single \emph{pair embedding}. Therefore, each one of the $M$ modalities yields a distinct pair embedding. The $M$ pair embeddings are combined in the multi-modal fusion layer. Finally, an output layer is learned to link the multi-modal pair embedding to the final similarity score. 

The \simcnn structure can therefore be trained independently to output a likelihood that a given pair is a true pair and then be used as a feature in addition to the other pre-computed \textit{Baseline features} in a logistic regression. Or, similarly to the Wide \& Deep architecture used in \citet{cheng2016wide}, we propose to learn \simcnn jointly with the logistic regression weights of the \textit{Baseline features}: we call this the \jsimcnn structure.

\section{Experiments and Results}\label{sec:results}

For the \simcnn implementation, we fixed the different filter sizes to be (2, 3, 4, 5, 6, 10) and the number of filters per size at 40. The dropout probability was set at 0.5 for training and 0 for test.
Negative pairs were sampled from the set of candidate pairs not present in the list of ground truth pairs. Training a multi-modal \simcnn, where $M=4$ with sequences generated from URLs at different depths, resulted in learning richer pair embeddings. Regarding the merging of cookie embeddings, we found that using element wise multiplication for the fusion layer and concatenation for the multi-modal layer improved the results. 

\begin{table}
\centering
\begin{tabular}{l|c|c|c}
Method & Precision & Recall & F1 \\
\hline\hline
Baseline & 40.01 & 42.12 & 41.04 $\pm0.04$\\
Baseline + \simcnn & 41.20 & 44.01 & 42.56 $\pm0.05$\\
\jsimcnn & 44.81 & 47.86 & \textbf{\textit{46.28} $\pm0.03$}\\
Baseline + \simcnn + XGB & 43.85 & 46.85 & 45.30 $\pm0.05$ \\
\jsimcnn with XGB & 44.83 & 47.89 & \textbf{\textit{46.79} $\pm0.03$}\\
\hline
\citet{Phan2017} & 39.30 & 55.10 & 45.90 \\
\end{tabular}
\caption{Note that these results are computed using only 50\% of the correct pairs as in the CIKM cup setup. Training the \simcnn structure jointly with other features is the most efficient method. Here, Baseline refers to the Baseline features defined previously.}
\label{final_results}
\end{table}

We compare the performance of the \simcnn feature with \jsimcnn using only 50\% of the test pairs, following the CIKM challenge setup. We report confidence intervals on the F1 score obtained by doing 50 random splits of the test pairs. Note that this is different to how \citet{Phan2017} reported their results, as they evaluated on the full test set.

Our main results are presented in Table~\ref{final_results}. We show that using all features in joint training is the most efficient method: \jsimcnn is doing better than the baseline features combined with \simcnn. Our best score was obtained when adding a gradient boosted feature (XGB), which was obtained by training an XGBoost model on the baseline features and using its prediction as a feature. We improved the state-of-the-art results of \citet{Phan2017} by 1.9\%.

\section{Conclusion}\label{sec:conclusion}
This paper adds to the literature showing that applying Natural Language Processing techniques to sequences can be very efficient, especially in the field of similarity learning. We introduced the \simcnn structure showing that siamese CNNs can be applied successfully to learn sequence similarity on non textual data. Furthermore, with the \jsimcnn architecture, we showed that learning the \simcnn representation jointly with other features brings additional value and beats state-of-the-art results.

These promising results were obtained while keeping to state-of-the-art methods for candidate generation. Besides, when working on online users, training a sequence representation such as \emph{doc2vec} for all cookies might not be realistic (e.g. billions of sequences).
In that case it could be helpful to use a \simcnn-like structure to infer sequence embeddings and perform approximate kNN search. Future work should also better integrate the temporal information (time gaps, daily and weekly patterns,...) in our cookie/pair representation.
A possibility could be to integrate the time-gaps in \emph{doc2vec} as additional information as in \citet{Vasile2016}, or to use RNNs as proposed by \citet{Smirnova2017}. Further work should investigate the use of RNNs for the task of similarity learning.

More generally, this structure developed for cookie matching can be generalized to any kind of sequence similarity task. For example, we plan to test our similarity learning structure on textual data for the SemEval-2018 workshop on Semantic Evaluation \footnote{SemEval-2018:~\url{http://alt.qcri.org/semeval2018/}}.

\bibliography{CRTO-XDevice-paper}
\end{document}